\documentclass[12pt,a4]{article}
\usepackage[dvips]{color,graphicx}
\usepackage{bm,psfrag,afterpage,color,amsmath, amssymb}

\makeatletter
\def\eqnarray{\stepcounter {equation}\let \@currentlabel =\theequation
\global \@eqnswtrue
\global \@eqcnt \z@ \tabskip \@centering \let \\=\@eqncr
$$\halign to \displaywidth \bgroup \@eqnsel \hskip \@centering
$\displaystyle \tabskip \z@ {##}$&\global \@eqcnt \@ne \hfil
${\mbox{}##\mbox{}}$\hfil &\global \@eqcnt \tw@
$\displaystyle \tabskip \z@ {##}$\hfil \tabskip \@centering
&\llap {##}\tabskip \z@ \cr}
\makeatother

\begin{document}

{\baselineskip = 8mm 

\begin{center}
\textbf{\LARGE Semi-supervised logistic discrimination for functional data} \\[5mm]
\end{center}

\begin{center}
{\large Shuichi Kawano$^1$ \ and \ Sadanori Konishi$^2$}
\end{center}

\begin{center}
\begin{minipage}{14cm}
{
\begin{center}
{\it {\footnotesize 
$^1$ Department of Mathematical Sciences, Graduate School of Engineering, \\ Osaka Prefecture University, 
1-1 Gakuen-cho, Sakai, Osaka 599-8531, Japan. \\

\vspace{1.2mm}

$^2$ Department of Mathematics, Faculty of Science and Engineering, Chuo University,\\
1-13-27 Kasuga, Bunkyo-ku, Tokyo 112-8551, Japan. \\
}}

\vspace{2mm}

skawano@ms.osakafu-u.ac.jp \hspace{5mm} konishi@math.chuo-u.ac.jp

\end{center}

\vspace{1mm} 

{\small {\bf Abstract:} \
Multi-class classification methods based on both labeled and unlabeled functional data sets are discussed. 
We present a semi-supervised logistic model for classification in the context of functional data analysis. 
Unknown parameters in our proposed model are estimated by regularization with the help of EM algorithm. 
A crucial point in the modeling procedure is the choice of a regularization parameter involved in the semi-supervised functional logistic model. 
In order to select the adjusted parameter, we introduce model selection criteria from information-theoretic and Bayesian viewpoints. 
Monte Carlo simulations and a real data analysis are given to examine the effectiveness of our proposed modeling strategy. 
}

\vspace{3mm}



}
\end{minipage}
\end{center}

\baselineskip = 8mm

\vspace{8mm}

\noindent {\Large \textbf{1 \ Introduction}}

\vspace{5mm}

\noindent In recent years, functional data analysis has been used in various fields of study such as chemometrics and meteorology (e.g., we refer to Ramsay and Silverman, 2002; 2005, Ferraty and Vieu, 2006). 
The basic idea behind functional data analysis is to express a discrete data set as a smooth function data set, and then exploit information obtained from the set of functional data using the functional analogs of classical multivariate statistical tools. 
Till this day, several researchers have studied a variety of functional versions of traditional supervised and unsupervised statistical methods; e.g., functional regression analysis (James and Silverman, 2005; Yao {\it et al.}, 2005; Araki {\it et al.}, 2009a), functional discriminant analysis (Ferraty and Vieu, 2003; Rossi and Villa, 2006; Araki {\it et al.}, 2009b), functional principal component analysis (Rice and Silverman, 1991; Siverman, 1996; Yao and Lee, 2006) and functional clustering (Abraham {\it et al.}, 2003; Rossi {\it et al.}, 2004; Chiou and Li, 2007).

Meanwhile, a semi-supervised learning, which is a modeling procedure based on both labeled and unlabeled data, has received considerable attention in the contemporary statistics, machine learning and computer science (see, e.g., Chapelle {\it et al}., 2006; Liang {\it et al.}, 2007; Zhu, 2008). 
In particular, it is known that the semi-supervised learning is useful in the application areas including text mining and bioinformatics, in which obtaining labeled data  is difficult while unlabeled data can be easily obtained. 
Many of ordinary statistical multivariate analyses have been extended into the semi-supervised resemblances by earlier researchers; e.g., semi-supervised regression analysis (Verbeek and Vlassis, 2006; Lafferty and Wasserman, 2007; Ng {\it et al.}, 2007), semi-supervised discriminant analysis (Miller and Uyer, 1997; Yu {\it et al.}, 2004; Zhou {\it et al}., 2004; Dean {\it et al.}, 2006; Kawano and Konishi, 2011) and semi-supervised clustering (Basu {\it et al.}, 2004; Zhong, 2006; Kulis {\it et al.}, 2009).

In this paper, our aim is to extend the supervised modeling procedures for functional data into semi-supervised counterparts. 
We, in particular,  focus on a multi-class classification or discriminant problem, and develop a semi-supervised logistic model for functional classification problem. 
Unknown parameters in the model are estimated by the regularization method along with the technique of EM algorithm. 
A crucial issue for the modeling procedure is to choose a value of a regularization parameter involved in the semi-supervised functional logistic model. 
In order to select the optimal value of the regularization parameter, we then introduce model selection criteria based on information-theoretic and Bayesian approaches that evaluate semi-supervised functional logistic models estimated by the regularization method. 
Some numerical examples including a microarray data analysis are illustrated to investigate the effectiveness of our modeling strategy.

This paper is organized as follows. 
In Section 2, we consider a functionalization method that converts the discrete data into the functional form using basis expansions. 
Section 3 proposes a functional logistic model in the context of the semi-supervised multi-class classification problem. 
In this section, we also present an estimation procedure based on the regularization method with the help of EM algorithm. 
Section 4 derives model selection criteria to select a regularization parameter in the functional logistic models. 
In Section 5, Monte Carlo simulations and a real data analysis are given to assess the performances of the proposed semi-supervised functional logistic discrimination. 
Some concluding remarks are given in Section 6.

\vspace{8mm}
\noindent {\Large \textbf{2 \ Functionalization}}

\vspace{5mm}

\noindent Suppose that we have $n$ independent observations $\mbox{\boldmath $x$}_1, \ldots,\mbox{\boldmath $x$}_n$, where $\mbox{\boldmath $x$}_{\alpha}$ consist of the $N_{\alpha}$ observed values $x_{\alpha 1},\ldots,x_{\alpha N_{\alpha}}$ at discrete times $t_{\alpha 1}, \ldots, t_{\alpha N_{\alpha}}$, respectively. 
Our aim in this section is to express a data set $\{ (x_{\alpha i},$ $t_{\alpha i}); $ $i=1,\ldots,N_{\alpha}, $ $~t_{\alpha i}\in \mathcal{T} \subset \mathbb{R}\}$ $(\alpha=1,\ldots, n)$ as a set of smooth functions $\{x_{\alpha}(t);\alpha=1,\ldots,n,$$~t\in\mathcal{T} \}$ by a smoothing technique. 
In this section we drop the notation on the subject $\mbox{\boldmath $x$}_\alpha$, and hence consider  a functionalization procedure of the data set $\{(x_i, t_i); $ $i =1, \ldots, N\}$.

It is assumed that the observed values $\{ (x_{i}, t_{i}); $ $i = 1,\ldots, N \}$ for a subject are drawn from a regression model as follows: 
\begin{eqnarray}
x_{i}=u(t_{i})+\varepsilon_{i}, ~~i = 1, \ldots, N_{},
\label{chF1}
\end{eqnarray}
where $u(t)$ is a smooth function to be estimated and the errors $\varepsilon_{i}$ are independently, normally distributed with mean zero and variance $\sigma^2$. 
We also assume that the function $u(t)$ can be represented by a linear combination of pre-prepared basis functions in the form
\begin{eqnarray}
u(t)=\sum_{k=1}^m \omega_{k} \phi_k(t; \mu_k, \eta_k^2),
\label{rbfeq1}
\end{eqnarray}
where $\omega_k$ are coefficient parameters, $m$ is the number of basis functions and $\phi_k(t; \mu_k, \eta_k^2)$ are Gaussian basis functions given by 
\begin{eqnarray}
\phi_k(t; \mu_k, \eta^2_k)=\exp\left\{-\frac{(t - \mu_k)^2} {2 \eta^2_k}
\right\}, \quad k=1,\ldots,m.  \label{rbfeq2}
\end{eqnarray}
Here $\mu_k$ are the centers of the basis functions and $\eta_k$ are the dispersion parameters.
In particular, we use Gaussian basis functions proposed by Kawano and Konishi (2007), and hence the centers $\mu_k$ and the dispersion parameters $\eta_k$ are determined as follows: for equally spaced knots $\tau_k$ so that $\tau_1 < \cdots < \tau_4 = \min (t) < \cdots < \tau_{m+1} = \max (t) < \cdots < \tau_{m+4}$, we set the centers and the dispersion parameters as $\hat{\mu}_k = \tau_{k+2}$ and $\hat{\eta} \equiv \hat{\eta}_k = (\tau_{k+2} - \tau_k)/3$ for $k=1,\ldots,m$, respectively. 
For details of the procedure, we refer to Kawano and Konishi (2007).

It follows that the nonlinear regression model based on the Gaussian basis functions can be written as 
\begin{eqnarray}
f(x_{i}|t_{i}; \mbox{\boldmath $\omega$}, \sigma^2) =\frac{1}{\sqrt{2\pi\sigma^{2}}}\exp\left[ -\frac{\left\{x_{i}-\mbox{\boldmath $\omega$}^T \mbox{\boldmath $\phi$}(t_{i}) \right\}^2}{2\sigma^2}\right], \quad i = 1,  \ldots, N,  
\label{rbf-Gau-mod}
\end{eqnarray}
where  $\mbox{\boldmath $\omega$} =(\omega_{1},\ldots,\omega_{m})^T$ and $\mbox{\boldmath $\phi$}(t) = (\phi_1(t), \ldots, \phi_m(t))^T$. 
The parameters $\mbox{\boldmath $\omega$}$ and $\sigma^2$ are estimated by maximizing the regularized log-likelihood function in the form
\begin{eqnarray}
\ell_{\zeta}(\mbox{\boldmath $\omega$}, \sigma^2) &=& \sum_{i=1}^{N} \log
f(x_{i}|t_{i}; \mbox{\boldmath $\omega$}, \sigma^2) -\frac{N \zeta}{2}%
\mbox{\boldmath $\omega$}^T {\mathcal K} \mbox{\boldmath $\omega$}  \nonumber \\
&=& -\frac{N}{2}\log(2\pi\sigma^2)-\frac{1}{2\sigma^2} (\mbox{\boldmath $x$}%
-\Phi\mbox{\boldmath $\omega$})^T(\mbox{\boldmath $x$}-\Phi\mbox{\boldmath
$\omega$})-\frac{N\zeta}{2} \mbox{\boldmath $\omega$}^T {\mathcal K} \mbox{\boldmath
$\omega$},  \label{rbf-pen-lik}
\end{eqnarray}
where $\mbox{\boldmath $x$}=(x_1,\ldots,x_N)^T$, $\Phi=(\mbox{\boldmath $\phi$}(t_1),\ldots, \mbox{\boldmath $\phi$}(t_N))^T$, $\zeta \ (>0)$ is a smoothing parameter and ${\mathcal K}$ is a positive semi-definite matrix defined by ${\mathcal K}=D_2^T D_2$, where $D_2$ is a second-order difference term. 
The regularized maximum likelihood estimates are given by 
\begin{eqnarray}
\hat{\mbox{\boldmath $\omega$}} =(\Phi^T\Phi + N \zeta \hat{\sigma}^2
{\mathcal  K})^{-1} \Phi^T\mbox{\boldmath $x$}, ~~~~~~~~ \hat{\sigma}^2 =\frac{1}{N}%
\sum_{i=1}^{N} \left\{x_{i} -\hat{\mbox{\boldmath $\omega$}}^T%
\mbox{\boldmath $\phi$}(t_{i})\right\}^2.  \label{rbf-reg-est}
\end{eqnarray}



We obtain the optimal number of basis functions $m$ and the value of the smoothing parameter $\zeta$ by using a model selection criterion GIC (Ando {\it et al.}, 2008) for each smooth curve as the minimizer of the form
\begin{eqnarray}
\mathrm{GIC} (\zeta) =N\log(2\pi\hat{\sigma}^2)+N+2\mbox{tr}\{QR^{-1}\},
\label{gicfst}
\end{eqnarray}
where $\hat{\sigma}^2$ is given in Equation (\ref{rbf-reg-est}) and the $m \times m$ matrices $Q$ and $R$ are, respectively, given by 
\begin{eqnarray}
Q &=& \frac{1}{N\hat{\sigma}^2}\left( 
\begin{array}{cc}
{\displaystyle \frac{1}{\hat{\sigma}^2}\Phi^T\Lambda^2\Phi-\zeta {\mathcal K} \hat{\mbox{\boldmath
$\omega$}}\mathbf{1}_N^T\Lambda \Phi} & {\displaystyle \frac{1}{2\hat{\sigma}^4}%
\Phi^T\Lambda^3\mathbf{1}_N- \frac{1}{2\hat{\sigma}^2}\Phi^T\Lambda\mathbf{1}%
_N} \\ 
{\displaystyle \frac{1}{2\hat{\sigma}^4}\mathbf{1}_N^T\Lambda^3\Phi- \frac{1}{2\hat{\sigma}%
^2}\mathbf{1}_N^T\Lambda\Phi} & {\displaystyle \frac{1}{4\hat{\sigma}^6}\mathbf{1}%
_N^T\Lambda^4\mathbf{1}_N- \frac{N}{4\hat{\sigma}^2}}%
\end{array}%
\right), \\
R&=&\frac{1}{N\hat{\sigma}^2}\left( 
\begin{array}{cc}
{\displaystyle \Phi^T\Phi+N\zeta\hat{\sigma}^2 {\mathcal K}} & {\displaystyle \frac{1}{\hat{\sigma}^2}\Phi^T\Lambda%
\mathbf{1}_N} \\ 
{\displaystyle \frac{1}{\hat{\sigma}^2}\mathbf{1}_N^T\Lambda\Phi} & {\displaystyle \frac{N}{2\hat{\sigma}^2}}%
\end{array}
\right),
\end{eqnarray}
where $\mathbf{1}_N=(1,\ldots,1)^T$ and $\Lambda= {\rm diag} \left[ x_1-\hat{\mbox{\boldmath $\omega$}}^T\mbox{\boldmath $\phi$}(t_1),\ldots,x_N-\hat{\mbox{\boldmath $\omega$}}^T\mbox{\boldmath $\phi$}(t_N) \right]$.

Hence, the observed discrete data $\{(x_{\alpha i},t_{\alpha i}); t_{\alpha i} \in \mathcal{T}, i=1, \ldots, N_\alpha\}$ $(\alpha =1, \ldots, n)$ are smoothed by the methodology described above, and we obtain a functional data set  $\{x_{\alpha}(t); \ \alpha=1,\ldots,n\}$  given by 
\begin{eqnarray}
\hat{u}(t)= \sum_{k=1}^m \hat{\omega}_{\alpha
k}\phi_k(t)\equiv x_\alpha (t), \qquad t \in\mathcal{T}.  
\label{rbf-smo-fun}
\end{eqnarray}
Figure \ref{fig:smooth} shows a sketch of the functionalization using Gaussian basis functions. 
Circles represent observed discrete data, the below solid curves basis functions pre-prepared and the above solid line the estimated smooth curve. 
For details of the functionalization step in functional data analysis, we refer to Ramsay and Silverman (2005) or Araki {\it et al.} (2009a).

\begin{figure}[t]
\centering
\includegraphics[width=12cm,height=7cm]{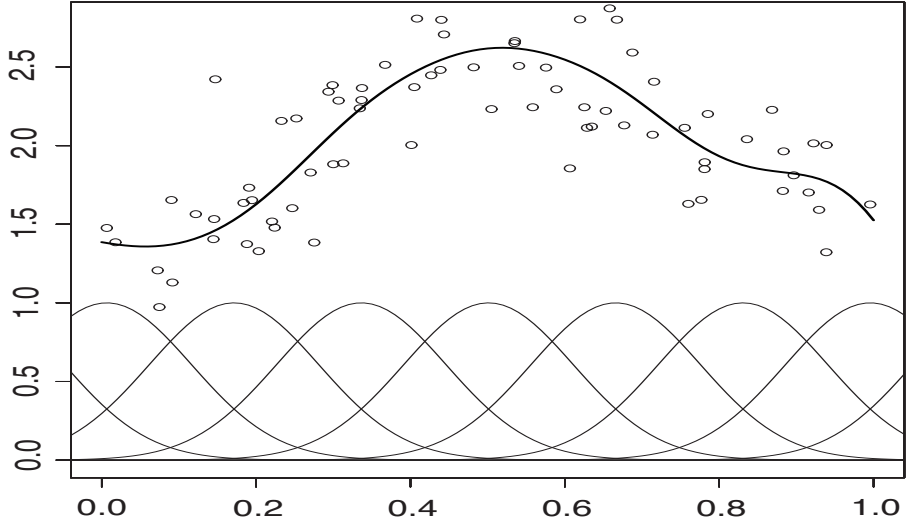}
\caption{Functionalization by Gaussian basis expansions}
\label{fig:smooth}
\end{figure}
%
%



\vspace{8mm}
\noindent {\Large \textbf{3 \ Semi-supervised functional logistic discrimination}}

\vspace{2mm}

\noindent {\large \textbf{3.1 \ Semi-supervised logistic model for functional data}}

\vspace{2mm}


\noindent In the framework of semi-supervised functional data analysis, we are given $n_1$ labeled functional data  $\{ (x_{\alpha} (t), g_{\alpha}) ; \alpha=1,\ldots,n_1,  \ t \in {\mathcal T}\}$ and $(n - n_1)$ unlabeled functional data $\{ x_{\alpha} (t) ; \alpha=n_1 + 1,\ldots,n,  \ t \in {\mathcal T} \}$. 
Here $x_{\alpha} (t)$ are functional predictors given in the previous section and $g_{\alpha} \in \{ 1, \ldots, L \}$ are group indicator variables in which $g=k$ implies that the functional predictor $x_{\alpha} (t)$ belongs to group $k$. 
First, a functional logistic model is constructed by using only labeled functional data $\{ (x_{\alpha} (t), g_{\alpha}) ; \alpha=1,\ldots,n_1, \  t \in {\mathcal T}\}$.

We consider the posterior probabilities for group $k \ (k=1,\ldots,L)$ given in a functional data $x_{\alpha} (t)$ as follows: ${\rm Pr} (g_{\alpha}=k | x_{\alpha})$. 
Under these posterior probabilities, Araki {\it et al.} (2009b) introduced a functional logistic model in the form
\begin{eqnarray}
{\rm log} \left\{ \frac{{\rm Pr} (g_{\alpha} = k | {x}_{\alpha} )}{{\rm Pr} (g_{\alpha} = L | {x}_{\alpha} )} \right\} = \beta_{kf} + \int x_{\alpha} (t) \beta_k (t) dt, \qquad k =1,\ldots,L-1.
\label{functional-logistic-model}
\end{eqnarray}
By using the same Gaussian basis function $\phi_j (t)$ as in Equation (\ref{rbfeq1}), $\beta_k (t)$ is assumed to be expanded as 
\begin{eqnarray}
\beta_k (t) = \sum_{j=1}^m \beta_{kj} \phi_j (t).
\label{beta-cof}
\end{eqnarray}
Then we can rewrite the functional logistic model in Equation (\ref{functional-logistic-model}) using the expansion in Equation (\ref{beta-cof}) as follows:
\begin{eqnarray}
{\rm log} \left\{ \frac{{\rm Pr} (g_{\alpha} = k | {x}_{\alpha} )}{{\rm Pr} (g_{\alpha} = L | {x}_{\alpha} )} \right\} = \beta_{kf} + \int x_{\alpha} (t) \beta_k (t) dt = {\bm \beta}_k^T {\bm z}_{\alpha},
\label{equation13}
\end{eqnarray}
where ${\bm \beta}_k = (\beta_{kf}, \beta_{k1}, \ldots, \beta_{km})^T$ and ${\bm z}_{\alpha} = (1, {\bm w}_{\alpha}^T J)^T$. 
Here $J$ is an $m \times m$ matrix with the $(i,j)$-th element
\begin{eqnarray}
J_{ij} = \sqrt{\pi \hat{\eta}^2} \exp \left\{ - \frac{(\hat{\mu}_i - \hat{\mu}_j)^2}{4 \hat{\eta}^2} \right\}, \qquad i,j = 1,\ldots,m,
\end{eqnarray}
where $\hat{\mu}_i$ and $\hat{\eta}$ are estimated centers and width parameters included in Gaussian basis functions in Section 2, respectively. 

Thus the conditional probabilities can be rewritten as
\begin{eqnarray}
{\rm Pr} (g_{\alpha} = k | x_{\alpha}) &=& \frac{\exp \{ {\bm \beta}_k^T {\bm z}_{\alpha} \}}{1 + \displaystyle{\sum_{j=1}^{L-1} \exp \{ {\bm \beta}_j^T {\bm z}_{\alpha} \}}}, \quad k=1,\ldots,L-1, \nonumber\\
{\rm Pr} (g_{\alpha} = L | x_{\alpha}) &=& \frac{1}{1+\displaystyle{\sum_{j=1}^{L-1} \exp \{ {\bm \beta}_j^T {\bm z}_{\alpha} \}}}.
\end{eqnarray}
We describe ${\rm Pr} (g_{\alpha}=k | x_{\alpha})$ as $\pi_k (x_{\alpha} ; {\bm \beta})$, since the probabilities depend on a parameter vector ${\bm \beta} = ( {\bm \beta}_1^T, \ldots, {\bm \beta}_{L-1}^T )^T$.

We introduce an $(L-1)$-dimensional response variable ${\bm y}_{\alpha} = (y_1^{(\alpha)},\ldots,y_{L-1}^{(\alpha)})^T \ (\alpha=1,\ldots,n_1)$, which indicates that  the $k$-th element of ${\bm y}_{\alpha}$ is set to 1 if the corresponding ${ x}_{\alpha} (t)$ belongs to the $k$-th class, for $n_1$ labeled functional data $\{ ({x}_{\alpha} (t), g_{\alpha}) ; \alpha=1,\ldots, n_1 \}$. 
Hence we obtain a multinomial distribution with the posterior probabilities $\pi_k ({ x}_{\alpha}; {\bm \beta})$ as follows:
\begin{eqnarray}
f({\bm y}_{\alpha} | {x}_{\alpha} ; {\bm \beta}) = \prod_{k=1}^{L-1} \pi_k ({x}_{\alpha} ; {\bm \beta})^{y_{k}^{(\alpha)}} \{ \pi_L ({x}_{\alpha} ; {\bm \beta}) \}^{1 - \sum_{j=1}^{L-1} y_j^{(\alpha)}}. 
\label{semi-supervised-linear-multinomial}
\end{eqnarray}

By introducing a dummy class label variable ${\bm t}_{\alpha}$ for unlabeled functional data $\{ x_{\alpha} (t) ; \alpha=n_1 + 1,\ldots,n \}$ given by 
\begin{eqnarray*}
{\bm t}_{\alpha} = (t_1^{(\alpha)},\ldots,t_{L-1}^{(\alpha)})^T = \left\{\begin{array}{ll}
(0,\ldots,0,1 \hspace{-0.7em}\raisebox{2.2ex}{\it\scriptsize{{\rm (}k{\rm )}}},0,\ldots,0)^{T} & \ {\rm if} \ \ {x}_{\alpha} (t) \ {\rm belongs \ to} \ k {\rm \mathchar"712D th \ class}, \vspace{3mm} \\
(0,\ldots,0)^T & \ {\rm if} \ \ {x}_{\alpha} (t) \ {\rm belongs \ to} \ L  {\rm \mathchar"712D th \ class}, \\
\end{array} \right.
\end{eqnarray*}
it is assumed that ${\bm t}_{\alpha}$ is distributed as the same multinomial distribution with the posterior probabilities $\pi_k (x_{\alpha} ; {\bm \beta})$ as in Equation (\ref{semi-supervised-linear-multinomial}). 
Also, for unlabeled functional data, we assume $\beta_{kf} + \int x_{\alpha}(t) \beta_k (t) = {\bm \beta}_k^T {\bm z}_\alpha \ (\alpha=n_1+1,\ldots, n; \ k=1,\ldots,L-1)$ similar to Equation (\ref{equation13}). 
The log-likelihood function based on both labeled and unlabeled functional data is then obtained by
\begin{eqnarray}
\ell ({\bm \beta}) &=& \sum_{\alpha = 1}^{n_1} \left[ \sum_{k=1}^{L-1} y_k^{(\alpha)} {\bm \beta}_k^T {\bm z}_{\alpha} - \log \left( 1 + \sum_{l=1}^{L-1} \exp \{ {\bm \beta}_l^T {\bm z}_{\alpha} \} \right) \right] \nonumber \\
&& {} + \sum_{\alpha = n_1+1}^{n} \left[ \sum_{k=1}^{L-1} t_k^{(\alpha)} {\bm \beta}_k^T {\bm z}_{\alpha} - \log \left( 1 + \sum_{l=1}^{L-1} \exp \{ {\bm \beta}_l^T {\bm z}_{\alpha} \} \right) \right].
\end{eqnarray}

\vspace{2mm}

\noindent {\large \textbf{3.2 \ Estimation via regularization}}

\vspace{2mm}

\noindent As mentioned in Araki {\it et al.} (2009b),  the maximum likelihood method often causes some ill-posed problems for a functional logistic model; i.e., unstable or infinite parameter estimates. 
Then we employ a regularization method to obtain the estimator of the parameters included in the functional logistic model. 
A regularization method achieves to maximize a regularized log-likelihood function
\begin{eqnarray}
\ell_{\lambda} ({\bm \beta}) = \ell ({\bm \beta}) - \frac{n_1 \lambda}{2} \sum_{k=1}^{L-1} {\bm \beta}^T_k K {\bm \beta}_k,
\label{FDA-regularized-function}
\end{eqnarray}
where $\lambda \ (>0)$ is a regularization parameter and $K$ is an $(m + 1) \times (m + 1)$ matrix given by  
\begin{equation}
K = \left(
\begin{array}{cc}
0 & {\bm 0}^T  \\
{\bm 0} & K^* \\
\end{array}
\right).
\end{equation}
Here ${\bm 0}$ is an $m$-dimensional zero  vector and $K^*$ is an $m \times m$ positive semi-definite matrix. 
In the section of numerical examples, we use an identity matrix as the matrix $K^*$. 

In maximizing the regularized log-likelihood function in Equation (\ref{FDA-regularized-function}), it is difficult to obtain the estimator of the parameters, since the values of dummy class labels ${\bm t}$ are unknown and $\partial \ell_{\lambda} ({\bm \beta}) / \partial {\bm \beta} = {\bm 0}$ does not have an explicit solution with respect to the parameter vector ${\bm \beta}$. 
Hence, we employ a following EM-based algorithm to obtain the estimator $\hat{\bm \beta}$. 
\begin{description}
\item[Step1] Initializing the parameter vector ${\bm \beta}$ by maximizing the regularized log-likelihood function via only labeled functional data $\{ ({x}_{\alpha} (t), g_{\alpha}) ; \alpha = 1,\ldots,n_1 \}$  with the help of Fisher's scoring method. 
\item[Step2] Construct a classification rule $\pi_k ({x}_{\alpha} ; \hat{\bm \beta})$. 
\item[Step3] By the use of the classification rule in Step2, compute the posterior probabilities $\pi_k ({x}_{\alpha}  ; \hat{\bm \beta}) \ (k=1,\ldots,L)$ for unlabeled functional data ${x}_{\alpha} (t) \ (\alpha = n_1+1,\ldots,n)$. 
According to the posterior probabilities, estimate ${\bm t}_{\alpha}$ as follows:  
\begin{eqnarray}
\hat{\bm t}_{\alpha} = (\hat{t}_1^{(\alpha)}, \ldots, \hat{t}_{L-1}^{(\alpha)})^T =(\pi_1 ({x}_{\alpha} ; \hat{\bm \beta}), \ldots, \pi_{L-1} ({x}_{\alpha} ; \hat{\bm \beta}))^T.
\end{eqnarray}
\item[Step4] Replace $t_k^{(\alpha)}$ into $\hat{t}_k^{(\alpha)}$ in the regularized log-likelihood function. 
Then estimate the parameter vector ${\bm \beta}$ using Fisher's scoring method. 
\item[Step5] Repeat the Step2 to the Step4 until the convergence condition
\begin{eqnarray}
| \ell_\lambda (\hat{\bm \beta}^{(k+1)}) - \ell_\lambda (\hat{\bm \beta}^{(k)}) | < 10^{-5}
\end{eqnarray}
is satisfied, where $\hat{\bm \beta}^{(k)}$ is the value of $\bm \beta$ after the $k$-th EM iteration. 
\end{description}

Therefore, we derive a statistical model $f ({\bm y} | x; \hat{\bm \beta})$ which is constructed by using both labeled and unlabeled functional data. 
The statistical model includes a tuning parameter; i.e., the regularization parameter $\lambda$. 
Since the selection of this parameter is regarded as the selection of candidate models, we introduce model selection criteria to choose the constructed models.

\vspace{8mm}
\noindent {\Large \textbf{4 \ Model selection criteria}}

\vspace{2mm}

\noindent 
In this section, we derive two types of model selection criteria to evaluate  semi-supervised functional logistic models from the viewpoints of information-theoretic and Bayesian approaches. 

\vspace{2mm}


\noindent {\large \textbf{4.1 \ Generalized information criterion}}

\vspace{2mm}

\noindent Akaike (1974) proposed the Akaike information criterion (AIC), which enables us to evaluate statistical models estimated by the maximum likelihood method. 
While the AIC is very useful for various fields of research, the criterion cannot be directly applied into models constructed by other estimation procedures.

Konishi and Kitagawa (1996) introduced an information criterion, which can evaluate models constructed by various estimation procedures including robust,  Bayesian  and regularization methods. 
Using the result of Konishi and Kitagawa (1996), we propose a generalized information criterion (GIC) in the context of the semi-supervised functional logistic model. 
The model selection criterion is given as follows:
\begin{eqnarray}
{\rm GIC} = -2 \sum_{\alpha=1}^{n_1} \log f ({\bm y}_{\alpha} | x_{\alpha}; \hat{\bm \beta}) + 2 {\rm tr} \left\{ Q (\hat{\bm \beta}) R^{-1} (\hat{\bm \beta}) \right\},
\end{eqnarray}
where the matrices $Q (\hat{\bm \beta})$ and $R (\hat{\bm \beta})$ are 
\begin{eqnarray}
Q (\hat{\bm \beta}) &=& \frac{1}{n_1} \left[ \{ (B-C) \odot A \}^T - \lambda E \hat{\bm \beta} {\bm 1}_{n_1}^T \right] \{ (B-C) \odot A \}, \\
R (\hat{\bm \beta}) &=& - \frac{1}{n_1} (C \odot A)^T (C \odot A) + \frac{1}{n_1} D + \lambda E,
\label{semi-functional-R}
\end{eqnarray}
with
\begin{eqnarray*}
A &=& (Z,\ldots,Z), \qquad n_1 \times (m+1)(L-1), \\
B &=& ({\bm y}_{(1)} {\bm 1}_{m+1}^T, \ldots, {\bm y}_{(L-1)} {\bm 1}_{m+1}^T)^T, \\
C &=& ({\bm \pi}_{(1)} {\bm 1}_{m+1}^T, \ldots, {\bm \pi}_{(L-1)} {\bm 1}_{m+1}^T)^T, \\
D &=& {\rm block \ diag} \{ Z^T {\rm diag} ({\bm \pi}_{(1)}) Z, \ldots, Z^T {\rm diag} ({\bm \pi}_{(L-1)}) Z \}, \\
E &=& {\rm block \ diag} (K, \ldots, K), \qquad (m+1)(L-1) \times (m+1)(L-1),\\
Z &=& ({\bm z}_1, \ldots, {\bm z}_{n_1})^T, \\
{\bm y}_{(k)} &=& (y_{k}^{(1)}, \ldots, y_{k}^{(n_1)})^T, \\
{\bm \pi}_{(k)} &=& (\pi_k (x_1 ; \hat{\bm \beta}), \ldots, \pi_k (x_{n_1} ; \hat{\bm \beta}))^T.
\end{eqnarray*}
Here the operator $\odot$ denotes the Hadamard product, which means the elementwise product of matrices; that is, $A_{ij} \odot B_{ij} = (a_{ij} b_{ij})$ for matrices $A_{ij}=(a_{ij})$ and $B_{ij}=(b_{ij})$.

\vspace{5mm}

\noindent {\large \textbf{4.2 \ Generalized Bayesian information criterion}}

\vspace{2mm}

\noindent In Bayesian inference, Schwarz (1978) presented the Bayesian  information criterion (BIC) from the viewpoint of maximizing a marginal likelihood. 
However, the BIC covers only models estimated by the maximum likelihood method.

By extending the Schwarz's (1978) idea, Konishi {\it et al.} (2004) derived a novel Bayesian information criterion to evaluate models estimated by regularization in the framework of generalized linear models. 
Hence, by using the result given in Konishi {\it et al.} (2004), we present a generalized Bayesian information criterion (GBIC) for evaluating the statistical model constructed by the semi-supervised functional logistic modeling procedure in the form
\begin{eqnarray}
{\rm GBIC} &=& -2 \sum_{\alpha=1}^{n_1} \log f ({\bm y}_{\alpha} | x_{\alpha}; \hat{\bm \beta}) + n_1\lambda \sum_{k=1}^{L-1} \hat{\bm \beta}_k^T K \hat{\bm \beta}_k - (L-1) \log |K|_+  \nonumber \\
{}{} && + \log |R (\hat{\bm \beta})| - (L-1) (m+1-d) \log \lambda - (L-1) d \log \left( \frac{2 \pi}{n_1} \right),
\end{eqnarray}
where $R(\hat{\bm \beta})$ is given by Equation (\ref{semi-functional-R}) and  $|K|_+$ is the product of the positive eigenvalues of $K$ with the rank $d$. 

We thus select a tuning parameter $\lambda$ by minimizing either the model selection criterion GIC or GBIC. 
For more details of derivations about the model selection criteria, we refer to Konishi and Kitagawa (2008).


\vspace{8mm}
\noindent {\Large \textbf{5 \ Numerical studies}}

\vspace{5mm}

\noindent We conducted some numerical examples to investigate the effectiveness of the proposed modeling procedure. 
Monte Carlo simulations and a real data analysis are given to illustrate our proposed semi-supervised functional modeling strategy.

\vspace{5mm}

\noindent {\large \textbf{5.1 \ Monte Carlo simulations}}

\vspace{2mm}

\noindent We demonstrated the efficiency of the proposed functional modeling procedure through Monte Carlo simulations. 
In the simulation study, we generated $n$ discrete samples $\{ (x_{\alpha t_{i}}, g_{\alpha}); \alpha =1,\ldots, n, \ i=1,\ldots, l \}$, where predictors $x_{\alpha t_i}$ are assumed to be obtained by $x_{\alpha t_i} = h_{\alpha} (t_i) + \varepsilon_{\alpha t_i}$ and the class label $g_{\alpha}$ indicates 1 or 2 which is the group number. 
We considered two settings as follows:
\begin{eqnarray*}
&&\mathrm{\bf{Case~1}}\\     
 &&h_{\alpha} (t_i) = \sin(c_{\alpha} t_i\pi) u_{\alpha},~ \varepsilon_{\alpha t_i}\sim
N(0,0.1),~t_i = \frac{2i-2}{49}, \ n=600, \ l=50,  \\
 &&g_{\alpha}=1 : c_{\alpha}=1, \ u_{\alpha} \sim U [0.3,1.3], \\ 
&&g_{\alpha}=2 : c_{\alpha}=1.02, \ u_{\alpha} \sim U[0.1,0.6],  
\end{eqnarray*}
\begin{eqnarray*}
&&\mathrm{\bf{Case~2}}\\     
 &&h_{\alpha} (t_i) = u_{\alpha} w (t_i) + (1 - u_{\alpha}) v (t_i), \ \varepsilon_{\alpha t_i} \sim N(0,1), \ t_i = \frac{i+4}{5}, \ n=600, \ l=101, \\
&&g_{\alpha}=1 : u_{\alpha} \sim U[0,1], \ w (t_i) = \max (6-|t_i-11|,0), \ v(t_i) = \max (6-|t_i-11|,0) - 4,\\
&&g_{\alpha}=2 : u_{\alpha} \sim U[0,1], \ w (t_i) = \max (6-|t_i-11|,0), \ v(t_i) = \max (6-|t_i-11|,0) + 4.
\end{eqnarray*}
Figure \ref{simulation-FDA} denotes the true functions $h (t)$ for the Cases 1 and the Case 2, respectively. 
We divided the data set into 300 training data and 300 test data with an equal prior probability for each class. 
In order to implement the semi-supervised method, the training data were randomly divided into two halves with labeled functional data and unlabeled functional data, where the labeled functional data were assigned as 5\%, 10\%, 20\%, 30\%, 40\%, 50\% and 60\% of the training data, respectively.

\begin{figure}[t]
\centering
\includegraphics[width=7.4cm,height=6cm]{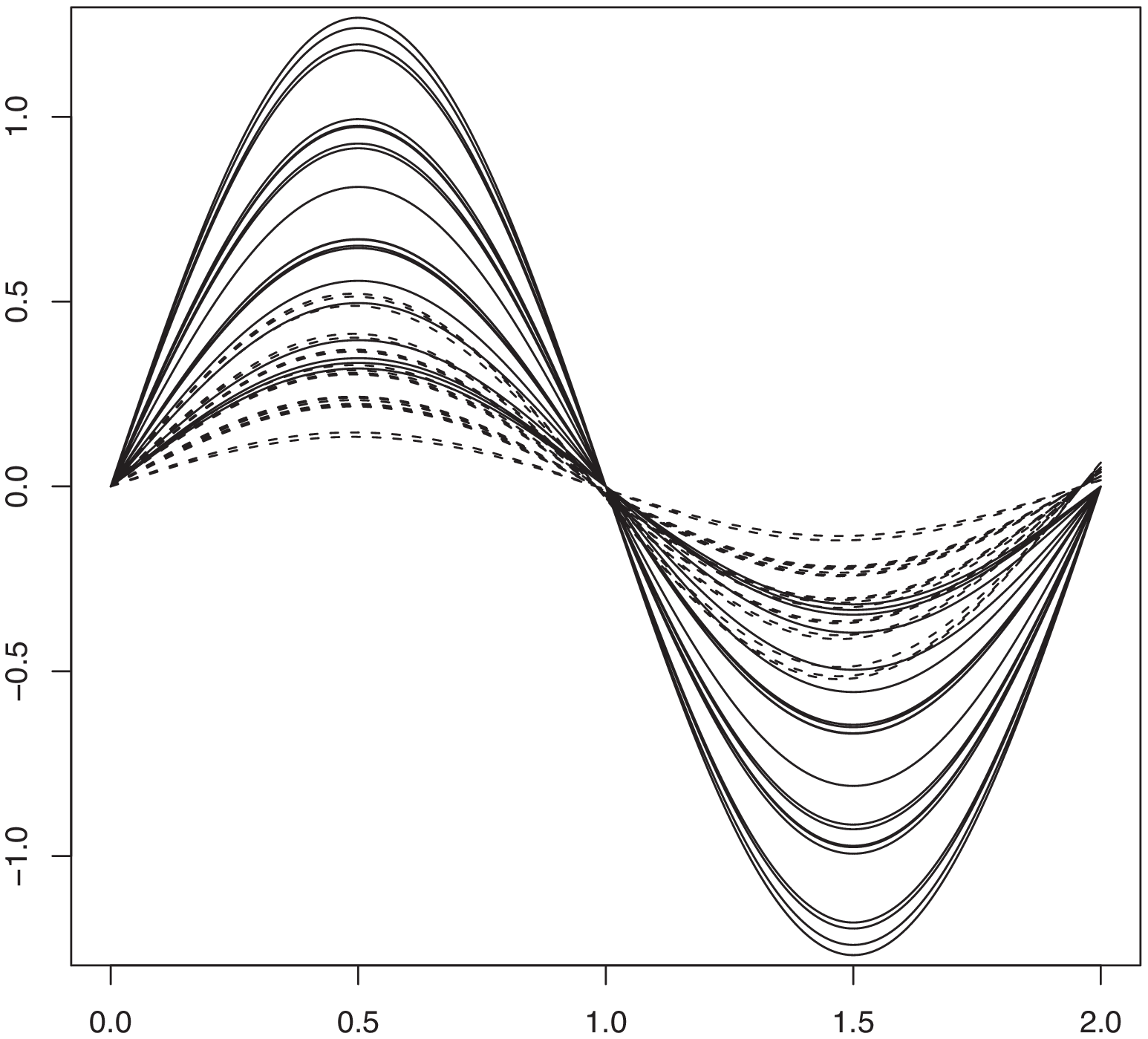}
\hspace{0.5cm} 
\includegraphics[width=7.8cm,height=6.8cm]{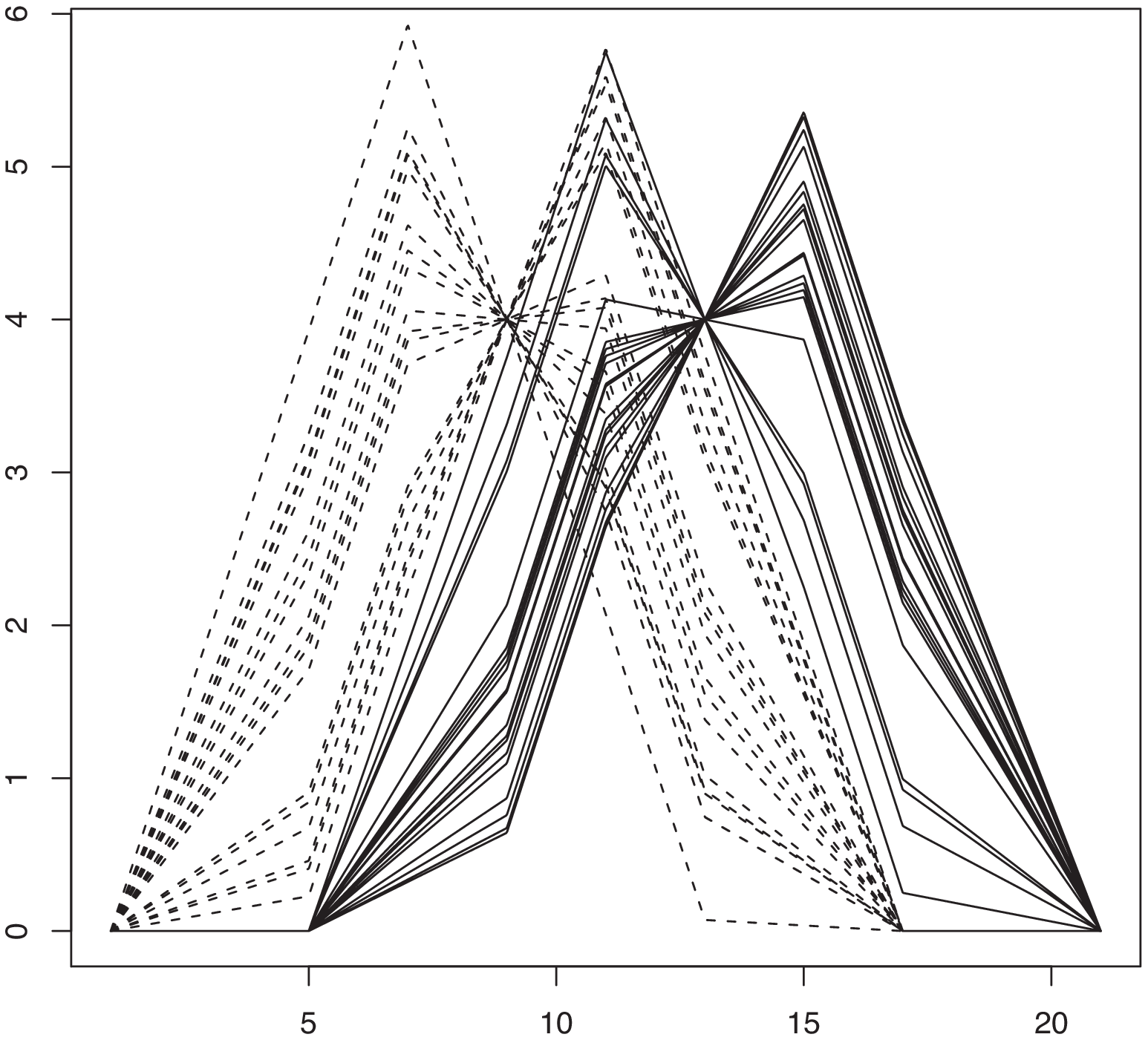} \\ \vspace{3mm}
\hspace{0cm} (a) \hspace{7cm} (b) 
\caption{True functions for (a) Case 1 and (b) Case 2. In each case, 
there are 10 subjects.
Solid lines represent the group 1, while dashed lines represent the group 2.}
\label{simulation-FDA}
\end{figure}

We compared the performances of semi-supervised functional logistic model (SFLDA) with those of supervised functional logistic model (FLDA) proposed by Araki {\it et al.} (2009b), support vector machine with the RBF kernel (SVM), $k$-nearest neighbor classification (KNN), functional support vector machine with the RBF kernel (FSVM) proposed by Rossi and Villa (2006), and semi-supervised methods proposed by Zhou {\it et al.} (2004) (LLGC: learning with local and global consistency) and Yu {\it et al.} (2004) (ILLGC: inductive learning with local and global consistency).  
The discrete data set was transformed into a functional data set using the smoothing technique described in Section 2. 
Semi-supervised and supervised functional modeling strategies  (i.e., SFLDA, FLDA and FSVM) were applied into the functional data set. 
The regularization parameter in the SFLDA and the FLDA was selected by using the GIC or the GBIC. 
For the GIC or the GBIC of the FLDA, we refer to Araki {\it et al.} (2009a; 2009b). 
Adjusted parameters included in the SVM, the FSVM, the LLGC and the ILLGC were optimized by the five-fold cross validation, respectively. 
The number of neighbors $k$ in the KNN was selected by the leave-one-out cross validation.

Tables \ref{FDA-simulation-testerror} and \ref{FDA-simulation-testerror2} show comparisons of the test error rates for the simulated data. 
These values were averaged over 50 repetitions. 
The average values of the tuning parameter $\lambda$ for 50 runs of the Case 1 were $\lambda=5.96 \times 10^{-5}$ for the GIC and $\lambda=9.48 \times 10^{-5}$ for the GBIC, while those of the Case 2 were $\lambda=1.00 \times 10^{-2}$ for the GIC and $\lambda=2.28 \times 10^{-2}$ for the GBIC. 
For the Case 1, we observe that the SFLDA methods evaluated by the GIC and the GBIC are superior to other methods except for the FLDA methods in almost all cases. 
Also, our proposed methods SFLDA seem to provide lower misclassification errors than the FLDA methods, when the size of labeled functional data is small (e.g., 10\% of training data). 
In the case of the Case 2, the SFLDA methods outperform the SVM, the KNN, the FSVM, the LLGC and the ILLGC in all situations with respect to minimizing the test errors. 
In addition, the proposed procedures SFLDA may be competitive or slightly superior to the FLDA methods. 


\begin{table}[t]
\begin{center}
\caption{Comparison of test errors with different percentages of labeled functional data in the training data set for the Case 1. 
Figures in parentheses indicate the model selection criteria used in the simulation study. }
\vspace{5mm}
\begin{tabular}{lcccccccc}
\hline
Method $\setminus$ \% & 5 & 10 & 20 & 30 & 40 & 50 & 60 \\ \hline
SFLAD (GIC)    &   0.269 & 0.210  & 0.202 &  0.192 &  0.189 &  0.186 & 0.185  \\
FLDA (GIC)   &   0.248 & 0.216 &  0.204 &  0.193 &  0.187 &  0.185 &  0.184 \\
SFLAD (GBIC)       & 0.271 &  0.210 &  0.202 &  0.193 &  0.188 &  0.185 &  0.185  \\
FLDA (GBIC)     & 0.359 & 0.237 &  0.200 &  0.188 &  0.185 &  0.183 &  0.182 \\
SVM &   0.278 & 0.221 &  0.203 &  0.195 &  0.194 &  0.183 &  0.185 \\
KNN     & 0.268 & 0.244 &  0.236 &  0.228&  0.225 &  0.220 &  0.215 \\
FSVM     & 0.322 & 0.266 &  0.253 &  0.231&  0.229 &  0.218 &  0.215 \\
LLGC     & 0.313 & 0.255 &  0.227 &  0.204&  0.197 &  0.192 &  0.187 \\
ILLGC    & 0.335 & 0.255 &  0.221 &  0.200&  0.193 &  0.189 &  0.185 \\
\hline 
\end{tabular}
\label{FDA-simulation-testerror}
\end{center}
\end{table}
\begin{table}[t]
\begin{center}
\caption{Comparison of test errors with different percentages of labeled functional data in the training data set for the Case 2. 
Figures in parentheses indicate the model selection criteria used in the simulation study. }
\vspace{5mm}
\begin{tabular}{lcccccccc}
\hline
Method $\setminus$ \% & 5 & 10 & 20 & 30 & 40 & 50 & 60 \\ \hline
SFLAD (GIC)     & 0.056 & 0.040 & 0.032  &  0.031 &  0.029 & 0.028 & 0.027 \\
FLDA (GIC)    & 0.056 & 0.043 & 0.035  &  0.029 &  0.029 &  0.029 & 0.027 \\
SFLAD (GBIC)      & 0.056 & 0.040 & 0.032 &  0.029 & 0.029 & 0.028 & 0.026 \\
FLDA (GBIC)    & 0.056 & 0.043 & 0.035  & 0.029 & 0.029 & 0.028 & 0.026 \\
SVM    & 0.075 & 0.056 & 0.040  & 0.037 & 0.034 & 0.030 & 0.031 \\
KNN    & 0.068 & 0.062 & 0.052  & 0.051 & 0.050 & 0.047 & 0.048 \\
FSVM    & 0.107 & 0.081 & 0.068  & 0.057 & 0.057 & 0.053 & 0.054 \\
LLGC    & 0.124 & 0.082 &  0.062 & 0.049 & 0.043 & 0.040 & 0.040 \\
ILLGC    & 0.111 & 0.049 & 0.040  & 0.035 & 0.031 & 0.030 & 0.030\\
\hline
\end{tabular}
\label{FDA-simulation-testerror2}
\end{center}
\end{table}


\vspace{5mm}

\noindent {\large \textbf{5.2 \ Microarray data analysis}}

\vspace{2mm}

\noindent We describe an application of the semi-supervised functional discriminant analysis to yeast gene expression data given in Spellman {\it et al}. (1998). 
This data set contains 77 microarrays and consists of two short time-courses (i.e., two time points) and four medium time-courses (18, 24, 17 and 14 time points). 
About 800 genes were classified into five different cell-cycle phases, namely, M/G1, G1, S, S/G2 and G2/M phases, while the other 5,378 genes were not classified. 
For more details of this data set, we refer to Spellman {\it et al}. (1998).

In our analysis, we used the ``cdc15-based experiment data'' sampled over 24 points after synchronization. 
For simplicity, any genes that contain missing values across any of the 24 time points were discarded. 
These expression data were considered to be a discretized realization of 632 expression curves evaluated at 24 time points. 
We functionalized the data using the smoothing methodology given in Section 2. 
A total of 300 genes were used as the training data set, and the remaining 332 genes were used as the test data set. 
We compared the SFLDA, which is our proposed semi-supervised functional method, with the FLDA, which is the supervised functional method.

First, we demonstrated the effectiveness of our semi-supervised methodology by setting functonal data with known class labels as unlabeled functional data. 
We randomly split the training data set into labeled functional data and unlabeled functional data, where 15\%, 20\%, 30\%, 40\% and 50\% of training data are allocated as labeled functional data, respectively, and we repeated the procedures 10 times. 
The values of the selected regularization parameter for 10 runs were $\lambda=2.80 \times 10^{-5}$ for the GIC and $\lambda=7.78 \times 10^{-4}$ for the GBIC. 
Figure \ref{semi-spellman1} shows the average precisions of the test data set for different ratios of labeled-unlabeled functional data in the training data set. 
On the $x$-axis, 15 means that 15\% of the training data was assigned as labeled functional data, and the remaining 85\% was used as unlabeled functional data. 
From the left panel of Figure 3, we observe that the SFLDA with the GIC seems to extract useful information from unlabeled functional data, since the SFLDA performs better than the FLDA in all cases. 
In contrast, the right panel of Figure 3 shows that the SFLDA is superior to the FLDA until 30\% labeled functional data, whereas the SFLDA is comparable to the FLDA in the range from 30\% to 50\% labeled functional data. 
\begin{figure}[t]
\centering
\includegraphics[width=7.4cm,height=6cm]{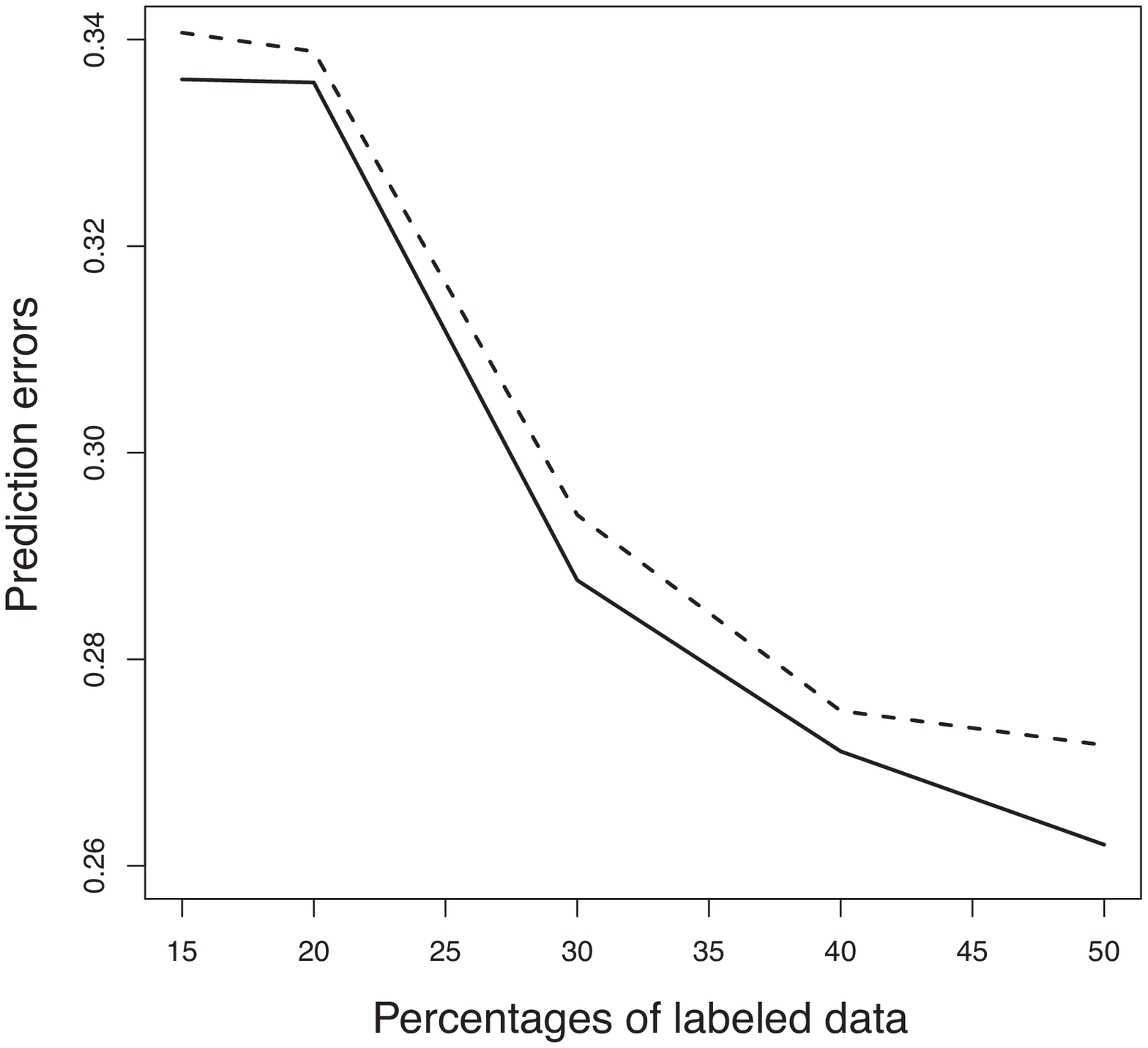}
\hspace{5mm}
\includegraphics[width=7.4cm,height=6cm]{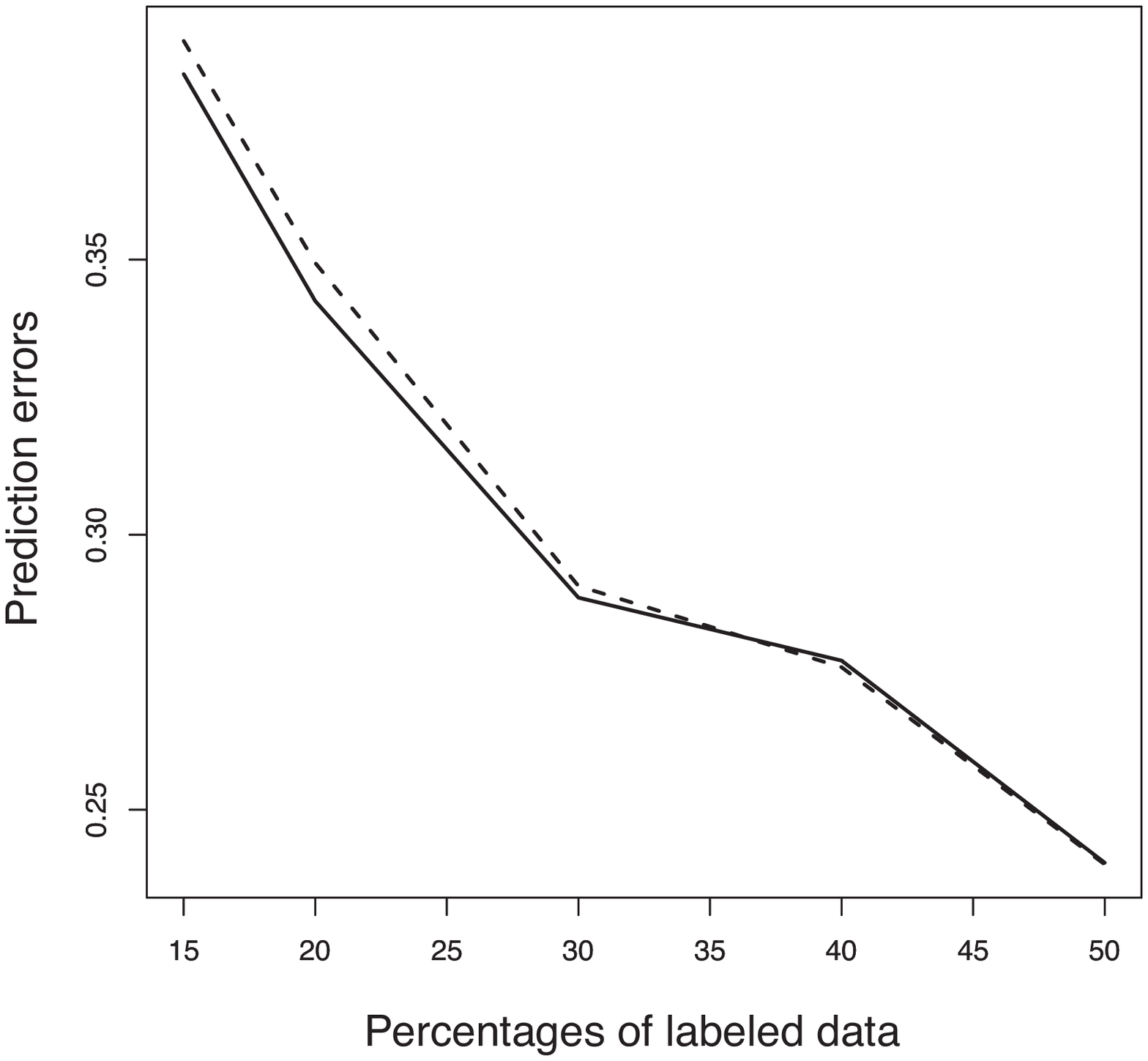} \\ \vspace{3mm}
\caption{Average prediction errors for several ratios of labeled functional data in the training data set. 
Solid line shows the result of the SFLDA while dashed line shows that of the FLDA. 
The left-hand panel indicates the results for the methods evaluated by the GIC, whereas the right-hand panel indicates those by the GBIC.}
\label{semi-spellman1}
\end{figure}
\begin{figure}[t]
\centering
\includegraphics[width=7.4cm,height=6cm]{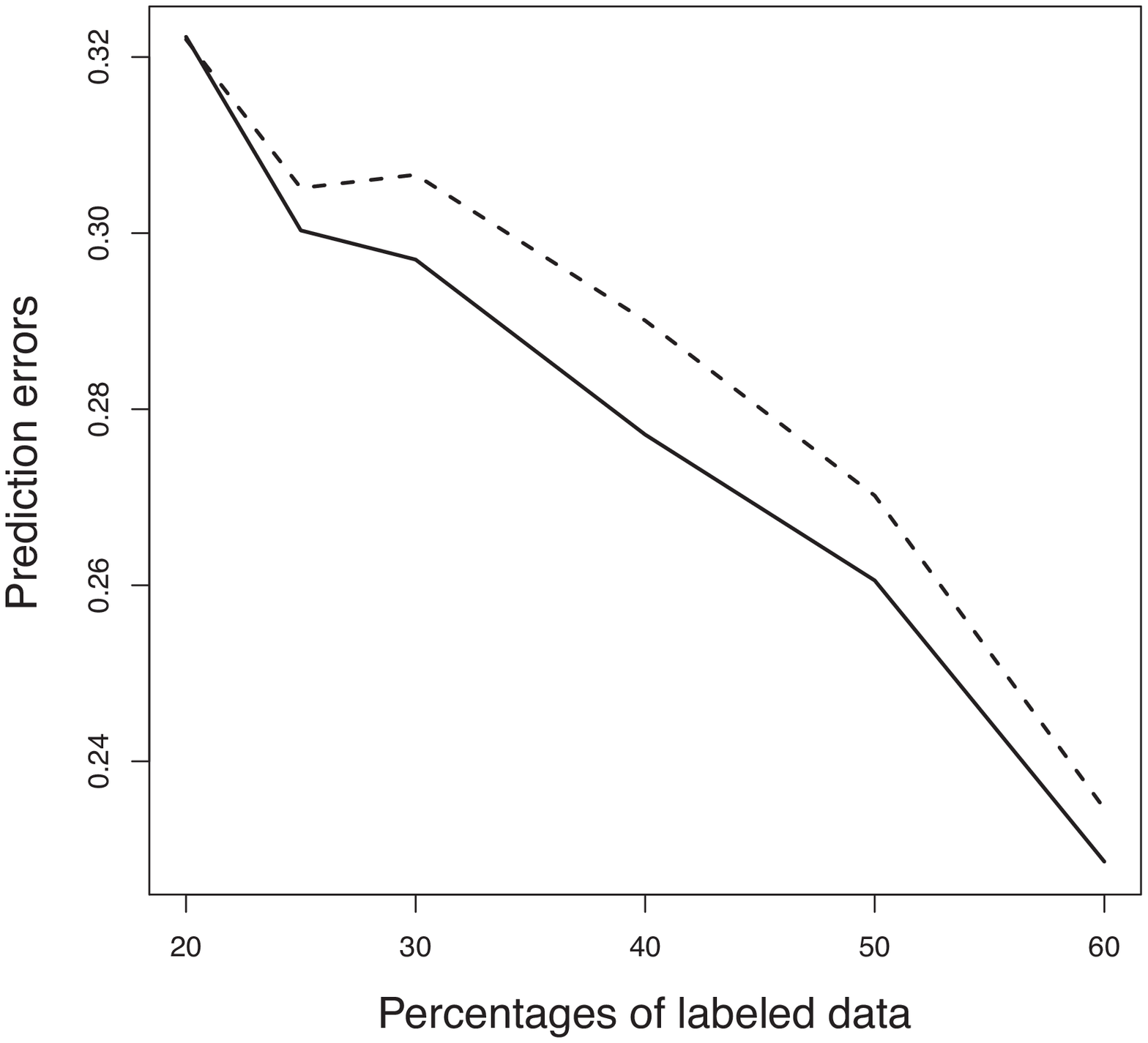}
\hspace{5mm}
\includegraphics[width=7.4cm,height=6cm]{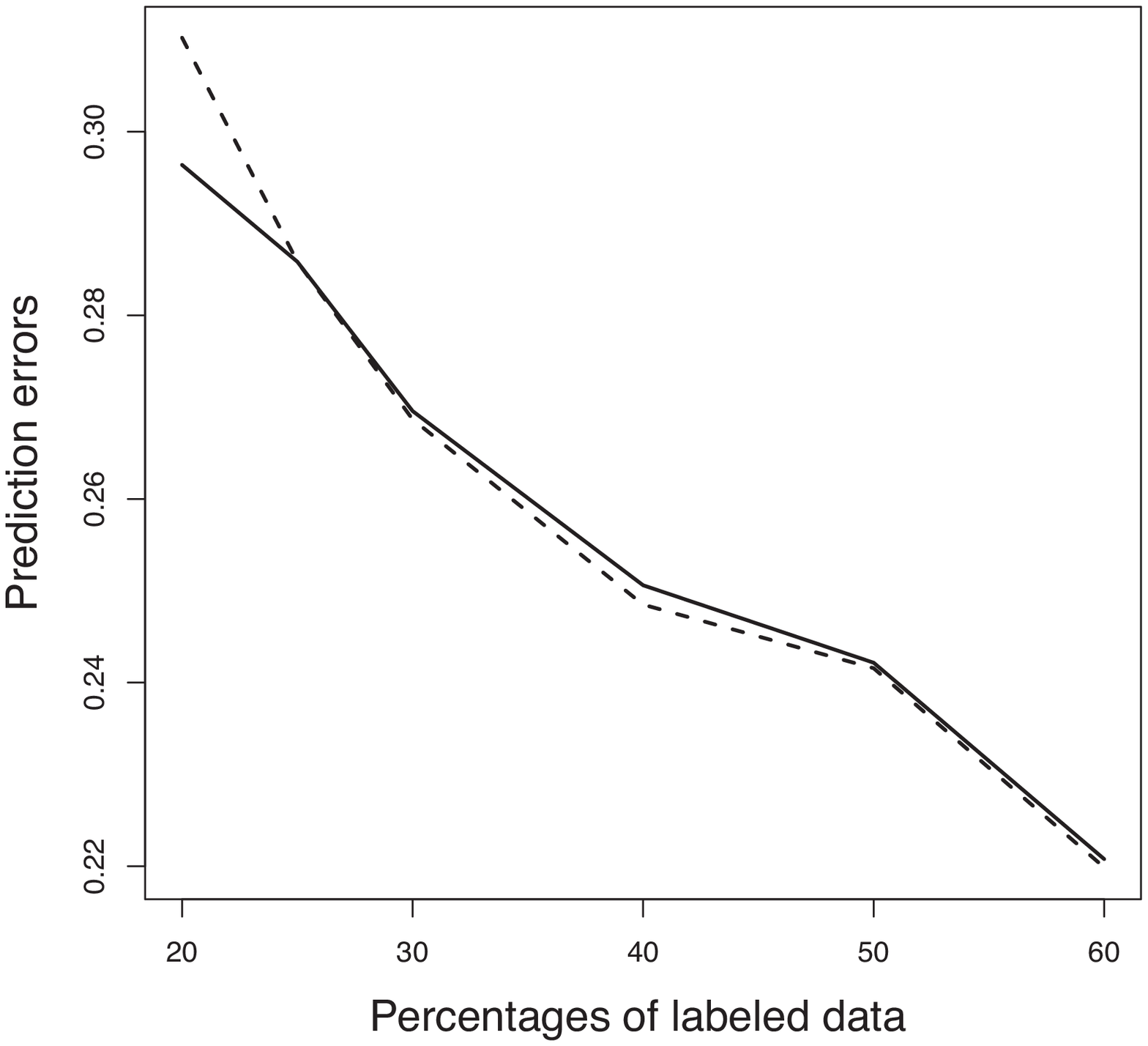} \\ \vspace{3mm}
\caption{Average prediction errors for several ratios of labeled functional data in the training data set, where we use real unlabeled functional data. 
Solid line shows the result of the SFLDA while dashed line shows that of the FLDA. 
The left-hand panel indicates the results for the methods evaluated by the GIC, whereas the right-hand panel indicates those by the GBIC.}
\label{real-unlabel-GBIC-GIC}
\end{figure}

Second, we examined the performances of our methods by using real unlabeled functional data which were not classified by Spellman {\it et al.} (1998). 
We prepared labeled functional data which consist of 20\%, 25\%, 30\%, 40\%, 50\% and 60\% of the training data, while unlabeled functional data  are set to 500 samples randomly selected from 5,378 real unlabeled examples. 
Our proposed models and the supervised functional models were applied into the data set. 
We repeated these procedures 10 times. 
We obtained the averaged optimal values of the regularization parameter for 10 repetitions as  $\lambda=1.00 \times 10^{-5}$ for the GIC and $\lambda=7.85 \times 10^{-5}$ for the GBIC. 
Figure \ref{real-unlabel-GBIC-GIC} shows the average test error rates for various ratios of labeled functional data in the training data set. 
For the left-hand panel of Figure \ref{real-unlabel-GBIC-GIC}, the SFLDA outperforms the FLDA without 20\% labeled functional data, while the SFLDA gives lower prediction errors than the FLDA on 20\% labeled functional data. 
Hence, these results suggest that real unlabeled functional data included in Spellman's {\it et al.} (1998) data set may have a potential for improving a prediction accuracy of our functional logistic procedures. 


\vspace{8mm}
\noindent {\Large \textbf{6 \ Concluding remarks}}

\vspace{5mm}

\noindent We proposed a semi-supervised functional logistic modeling procedure for the multi-class classification problem with the help of regularization. 
On the step of functionalization, a smoothing method using Gaussian basis expansions was applied to the observed discrete data set. 
A crucial issue for our semi-supervised modeling process is the choice of the regularization parameter $\lambda$. 
In order to select the value of the parameter, we introduced model selection criteria from the viewpoints of information-theoretic and Bayesian approaches. 
Monte Carlo simulations and a microarray data analysis showed that our modeling strategy yields relatively lower prediction error rates than previously developed methods. 
A further research should be to construct a semi-supervised functional regression modeling or clustering.


\vspace{5mm}
\noindent \textbf{\large Acknowledgement}}\\
This work was supported by the Ministry of Education, Science, Sports and Culture, Grant-in-Aid for Young Scientists (B), $\#$24700280, 2012--2015.



\end{document}